\documentclass{PoS}

\title{A simple idea for Lattice QCD at finite density}

\ShortTitle{A simple idea for Lattice QCD at finite density}

\author{\speaker{Rajiv V.\ Gavai}\thanks{Sir J.C. Bose National Fellow.}\\
         Department of Theoretical Physics, Tata Institute of Fundamental
         Research,\\ Homi Bhabha Road, Mumbai 400005, India.  \\
        E-mail: \email{gavai@tifr.res.in}}

\author{Sayantan Sharma\\
        Brookhaven National Laboratory,\\
        Upton, New York 11973.\\
        E-mail: \email{sayantans@bnl.gov}}

\abstract{We pursue the idea of adding the na{\i}ve $\mu N$ term, where $\mu$ is
the quark chemical potential and $N$ is the conserved quark number, to the
lattice QCD action. While computations of higher order susceptibilities,
required for estimating the location of the QCD critical point, need a lot
fewer number of quark propagators at any order as a result, it has its problem.
We discuss a solution, and examine if it works.}

\FullConference{9th International Workshop on Critical Point and Onset of Deconfinement - CPOD2014,\\
		17-21 November 2014\\
		ZiF (Center of Interdisciplinary Research), University of Bielefeld, Germany}

\begin{document}

\section{Introduction}

QCD at finite density is one of the most fascinating topics of research
currently because of the tantalizing experimental possibility of exploring the
the possible region of QCD critical point in the $T$-$\mu$ phase diagram.  The
challenges of the conceptual difficulties, such as the complex fermion
determinant and getting the physics of axial anomaly correct on the lattice make
it further attractive.  A lot of progress has been made in recent
years~\cite{gertcpod} making it  thus a very active field of research. In the
early years of lattice QCD, one of the important problem was how to incorporate
a finite chemical potential, $\mu$, in the local Dirac operator on the lattice.
Naively introducing $\mu$ as a Lagrange multiplier of the local point-split 
conserved charge $\bar\psi_x\gamma_4\psi_{x+\hat 0} + \bar\psi_{x+\hat 0} 
\gamma_4\psi_x$ led to $\mu$-dependent divergences in the free energy
of quarks in the continuum limit. Arguing that since $\mu$ enters in the
continuum like a fourth component of a static gauge field, the chemical
potential should enter on the lattice as $\rm{e}^{\pm\mu a}$ factors,
multiplying the forward and backward temporal gauge links
respectively~\cite{hk}. It did eliminate the undesired divergences mentioned
above. However, this proposal is not unique though: weights $( 1 \pm a
\mu)/\sqrt ( 1- a^2 \mu^2)$ to the temporal links also lead to finite results
for the free energy~\cite{bilgav}.  In fact the most general weights $f(\mu
a),g(\mu a)$ should follow $ f(a\mu) \cdot g(a \mu) =1$ with $f(0) = f'(0) =1$
in order to cure the divergences on the lattice~\cite{gavai}.
The analytical proof of absence of divergences in all these cases above was for
free quarks. Indeed, further numerical computations in quenched QCD showed it to
work for the interacting case as well~\cite{ggquen}, while a similar check in
the full theory is still lacking.

We revisit the problem of introduction of chemical potential on
the lattice again with a view to develop a simpler alternative.
We foresee two important applications in which the alternative we pursue here
may be of practical relevance in the coming years. Heavy ion collision
experiments at RHIC at Brookhaven National Laboratory and at FAIR, GSI may
likely produce strongly interacting systems at finite density. An important
theoretical input for modeling of the evolution of these systems is the QCD
equation of state(EoS). In order to calculate the EoS as a function of $\mu$ one
needs higher order quark number susceptibilities. The latter are also important
measurables on the lattice which relate directly to the fluctuations of net
baryon number \cite{ggplb} that are being measured in the heavy ion collision
experiments. Furthermore, these are also important in estimating the location of
the critical point in the QCD phase diagram.  The critical point is governed by
the singularity of the Taylor expansion of the baryon number susceptibility in
$\mu_B/T$. Its location can be determined by the radius of convergence of this
Taylor expansion.  It, in turn,  depends on the ratios of the higher order quark
number susceptibilities. Therefore, their accurate and efficient measurement on
the lattice is necessary although a rather challenging problem.  Having a Dirac
operator with a linear $\mu N$ term, simplifies the problem quite a lot.  For
example, the number of the inversions of the Dirac operator needed to calculate
the eighth order quark number susceptibility with the naive staggered quarks
reduces from $20$ to $8$~\cite{gs10}. For other improved Dirac operators and even higher
order QNS, the gain is even more which leads to a significant reduction of the
computational time.

Since the study of QCD critical point would be a major problem of interest in
the coming years both from experimental and theory perspective, it is also
important to consider how crucial is the role of exact chiral fermions at finite
density for the lattice studies.  The presence of the critical end-point depends
crucially on the number of light quark flavours. Model-based considerations
favour its presence for QCD with only two light quarks~\cite{piswilc}.  In addition, it needs
the anomalous $U_A(1)$ symmetry-restoration to take place at sufficiently high
temperatures~\cite{piswilc}.  A moderately heavy strange quark may affect the location of the
QCD critical end-point quantitatively but not its existence. Majority of the
calculations on QCD at finite density on the lattice employ staggered fermions
and its improved versions.  However, the continuum flavour and spin symmetries
are intermingled for the staggered quarks and the flavour singlet $U_A(1)$
anomaly is recovered only in the continuum limit. As is well-known, the overlap~\cite{neunar} 
and/or domain wall fermions~\cite{kaplan} are much more preferable from the chiral symmetry
perspective. They have  both the correct chiral and flavour symmetry on lattice
as well as an index theorem on the lattice \cite{hln,Lues}.  These are likely to
be crucial for investigations of the QCD critical point.  At present the
computations with chiral fermions are prohibitively expensive.

However, recently domain wall fermions have been used to measure the chiral
crossover transition temperature on the lattice~\cite{dweos}. This gives us a hope that with
the increase in computational resources and smarter algorithms, use of chiral
fermions would be more realistic in the coming years. 
Non-locality of the overlap fermions makes the introduction of the chemical
potential nontrivial.  
Bloch and Wettig \cite{wettig} proposed to use the same exponential prescription
as above for the timelike links of the Wilson-Dirac kernel $D_W(a \mu)$ to
define the corresponding overlap Dirac matrix at nonzero density.  The free energy from this 
overlap fermion action has no $a^{-2}$ divergences \cite{gl,bgs} in the
free case.  Unfortunately, however, it has no chiral invariance for nonzero
$\mu$ either \cite{bgs}.  Using the definition of the chiral projectors for
overlap fermions, we \cite{gsov} proposed a chirally invariant Overlap action
for nonzero $\mu$ : 
\begin{eqnarray} 
\nonumber
S^F &=& \sum_n  [\bar \psi_{n,L} (aD_{ov} + a\mu
\gamma^4) \psi_{n,L}+\bar \psi_{n,R} (aD_{ov} + a\mu \gamma^4) \psi_{n,R}]  \\
\nonumber &=& \sum_n  \bar \psi_n [ aD_{ov} + a\mu \gamma^4
( 1- aD_{ov}/2) ] \psi_n  ~. 
\end{eqnarray} 

It was shown that the fermion action is invariant under the Luescher
transformation ~\cite{Lues} and the corresponding order parameter, the chiral
condensate  in unique for all values of $a \mu$ \cite{gsov}. Moreover the anomaly is
$\mu$-independent on the lattice as is expected from the continuum~\cite{ns}. It, however,
has the same $\mu$-dependent $a^{-2}$ divergences in the number density and the
energy density as the linear $\mu$-case for naive/staggered fermions \cite{ns}.
Furthermore, unlike that case, these cannot be removed  by
exponentiation of the $\mu$-term \cite{ns}.  So either one has exact chiral invariance on
the lattice or have to deal with the divergences in the continuum limit of $a
\to 0$.  One needs to understand the origin of these $\mu$-dependent divergences
better. In quantum field theories, removal of divergences from physical quantum
theories has lead to extensive studies of the different regulators and their
suitability for different physical problems. We aim to have an understanding of
the nature of the divergences at finite density, and look for methods to remove 
them in alternate, perhaps simplifying, ways.

\section{The divergences at finite density : Why and how to remove them }

\subsection{Non-interacting fermions at finite chemical potential}
\begin{figure}[htb]
\hskip 4.5 cm \includegraphics[scale=0.4]{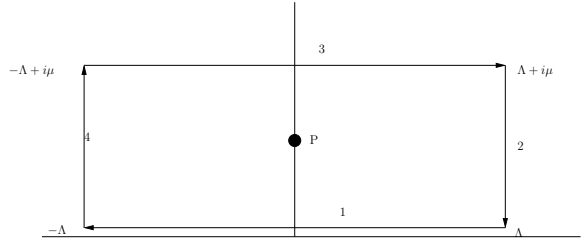}
\caption{The contour integral used for calculating the zero temperature part of the number density.}
\label{fig:cont}
\end{figure}

As a first step we examined carefully the free dense quark gas in continuum.  We
found that contrary to the common belief, these divergences are {\em not} due to
lattice artifacts.   Indeed, the $\mu$-dependent divergences exist in the
continuum theory as well when appropriate care is taken in manipulating
divergent integrals. The lattice regulator simply makes it easy to spot them.
Using a Pauli-Villars cut-off $\Lambda$ in the continuum theory, one can also
show the presence of $\mu \Lambda^2$ terms in the number density easily
\cite{gs14}.  We will sketch the argument here briefly, referring the reader to
\cite{gs14} for more details.  Let us recall that QCD 
partition function can be written in the path integral formalism as,
\begin{equation}
 \mathcal{Z}=\int \mathcal{D} A_\mu \mathcal{D}\bar \psi \mathcal{D} \psi
\rm{e}^{\int_0^{1/T} d\tau \int d^3 x\left[-1/2 Tr(F^2_{\mu,\nu})-\bar\psi
(\gamma_\mu(\partial_\mu-ig A_\mu)+m)\psi\right]}
\end{equation}
where ($\bar \psi$) $\psi$ and $A_\mu$ represents the (anti-)fermion and
the gluon fields respectively.  One introduces a chemical potential $\mu_i$ for
each conserved charge $\mathcal{N}_i  = \int d^3 x \bar \psi_i \gamma_4 \psi_i$,
leading to the additional term $\mu_i \mathcal{N}_i$ in the action above. 
Canonical definitions consisting of first and second derivative of $\mathcal{Z}$
with respect to $\mu_q=\mu$ yield quark number density and susceptibility,
\begin{equation}
 n=-\frac{T}{V}\frac{\partial \ln \mathcal{Z}(\mu)}{\partial \mu}|_{T=fixed} ~~~;
\qquad 
\chi=-\frac{T}{V}\frac{\partial^2 \ln \mathcal{Z}(\mu)}{\partial \mu^2}|_{T=fixed}
 \end{equation}
These can be calculated analytically when the gauge interactions
are switched off.  For simplicity we consider only massless fermions though this
derivation can be easily extended to finite quark mass.  The expression for
the free quark number density is then
\begin{equation}
\nonumber
 n=\frac{4iT}{ V}\sum_{n}\int \frac{d^3p}{(2\pi)^3}\frac{(\omega_n+i\mu )}
 {p^2+(\omega_n+i\mu )^2}~ \equiv \frac{4iT}{ V}\sum_{n}\int \frac{d^3p}{(2\pi)^3} \sum_{\omega_n}
 F(\omega_n,\mu ,\vec p),
\nonumber
\end{equation}
where $p^2= p_1^2+ p_2^2+ p_3^2$.  All the gamma matrices are Hermitian in our
convention.  $T=0$ and $\mu=0$ corresponds to the vacuum contribution which can
be removed by subtracting $n(T=0, \mu=0)$.  Although this is identically zero
for number density, the corresponding subtraction for the energy density is
actually $ \propto \Lambda^4$. 
Due to the Fermi-Dirac distribution functions, one does not expect any
ultraviolet divergences at finite $T$.  We therefore consider the $T=0$
contribution of the above expression to examine the presence of divergences, if
any.  The sum over $\omega_n$ turns into a continuous $\omega$ integral and the
explicit expression for the zero-$T$ contribution to the number density is,
\begin{equation}
\label{eqn:ncont}
n= 4i \int_{-\infty}^{\infty} \frac{d \omega}{2\pi} 
\frac{d^3p}{(2\pi)^3}\frac{(\omega+ i\mu )} {p^2+(\omega+ i\mu )^2}~.
\end{equation}
Under a variable transformation $\omega + i \mu = \omega'$, it can be recast as
\begin{equation}
\label{eqn:ncont1}
n= 4i\int_{-\infty+i\mu}^{\infty+i\mu} \frac{d \omega'}{2\pi} 
\frac{d^3p}{(2\pi)^3}\frac{\omega'} {p^2+\omega'^2}~.
\end{equation} 
In calculating this expression one considers a suitable contour as shown in
Figure ~\ref{fig:cont} with a cut-off $\Lambda$ for the $\omega$ integral in
order to regulate the divergent terms, and identify them by powers of $\Lambda$.
The expression in Eq.~\ref{eqn:ncont1} can be seen to be the line integral 3 in
Figure ~\ref{fig:cont}.  By adding and subtracting the line integrals labeled
as $1,2,4$, one obtains the result from the contour integral in terms of the
residue of the quark propagator pole with the corresponding $\theta$-function
defining the Fermi surface and leading to the usual $\mu^3$-term.  Out of the
subtracted terms,  the line integral $1$ is exactly zero due to CP symmetry.
However, the contribution of the arms 2 and 4 in the Figure ~\ref{fig:cont}
clearly do not cancel each other. The $\mu \Lambda^2$ terms can be shown to
arise from the algebraic sum of these terms \cite{gs14}, being of the form log
$\left[ \frac{p^2+(\Lambda+i \mu)^2}{p^2+(\Lambda-i \mu)^2}\right]$.  In the
limit $\Lambda \gg \mu$, and with the same cut-off $\Lambda$ for momentum $p$,
an expansion in $\mu/\Lambda$ shows that while the leading $\Lambda^3$ terms do
cancel from the numerator and denominator, the $\mu\Lambda^2$ terms add, and
survive. It is easy to see that a similar $\mu^2\Lambda^2$ divergent term
persists in the energy density as well.  It is amusing to note that the
divergence is absent in the number density for nonzero isospin chemical
potential with $\mu_B=0$, since this amounts to putting two flavours of quarks
in the Lagrangian with chemical potentials $\mu_I$ and $-\mu_I$. The divergent
term then cancels explicitly between the two quark flavours since it comes with
an equal and opposite sign.  The corresponding energy density, however, does
have the $\mu^2\Lambda^2$ divergent term.  On the other hand, the divergence
does exist for imaginary chemical potential for the number density as well,
as we show below.

\subsection{Non-interacting fermions in presence of imaginary $\mu$}

\label{sec:immu}
The expression for the number density for free fermions in presence of 
imaginary chemical potential at zero temperature is  
\begin{equation}
\label{eqn:ncontim}
n= -4i\int_{-\Lambda}^{\Lambda} \frac{d \omega}{2\pi} 
\frac{d^3p}{(2\pi)^3}\frac{(\omega-\mu )} {p^2+(\omega-\mu )^2}~.
\end{equation}
Performing the $\omega$ integral, which in this case is on real $\omega$-axis
only, one gets,
\begin{equation}
n= -4i \int\frac{d^3p}{(2\pi)^4}\ln\left(\frac{p^2+(\Lambda-\mu)^2} 
{p^2+(\Lambda+\mu)^2}\right )~.
\end{equation}
In the limit $\Lambda\gg \mu $, one can again expand the numerator and the 
denominator of the  integrand.   As above, we further assume the same
cut-off $\Lambda$ for the momentum integral.

Keeping terms upto $1/ \Lambda^3$, one obtains the number density 
as,
\begin{equation}
n=i\frac{\mu^3}{3\pi^2}+i\frac{\mu \Lambda^2}{4 \pi^2}~.
\end{equation}
The imaginary nature of number density comes out, as expected.  Indeed, one
recovers the {\em same~quadratic} divergence back if one perform analytic 
continuation to real $\mu$.  This is reassuring since it shows that
the presence of divergence  for nonzero $\mu$ is not due to the particular 
contour method we chose in Figure ~\ref{fig:cont}, as the integrals in this case are 
performed on the real $\omega$-axis for imaginary chemical potential.

\subsection{Free quark gas on the lattice}

We have seen that for non-interacting fermions within a cut-off regulator scheme
one obtains a $\mu\Lambda^2$ divergence. This explains the presence of a similar
divergence in the lattice expression of number density for non-interacting
fermions, since the lattice regulator is simply a
cut-off regulator and the $\mu/a^2$ term in the number density on a lattice is
just a manifestation of the continuum divergence.  This can be explicitly seen
following the same computation as above. The expression of
the number density on a $N^3\times N_T$ lattice with linear $\mu$ and a
conserved point-split charge is,
\begin{equation}
 \label{eqn:nodensitylm}
 na^3=\frac{i}{ N^3 N_T}\sum_{\vec p,n}\frac{(\sin\omega_n+i\mu a\cos\omega_n)\cos\omega_n}
 {f +(\sin\omega_n+i\mu a\cos\omega_n)^2}~.
\end{equation}
where $f=(ma)^2+\sin^2(a p_1)+\sin^2(a p_2)+\sin^2(a p_3), p_j=2 n_j \pi/N,~ n_j
\in Z$. 
At  $T=0$ one converts the sum over the Matsubara frequencies into an integral,
and obtains a line integral in the complex $\omega$ plane analogous to the line
integral marked $3$ in Figure ~\ref{fig:cont}.  It can be calculated using a
contour as in Figure ~\ref{fig:cont} but with $\tanh^{-1}(a\mu)$ in place of $\mu$.
Again the residue of the contour integral yields the Fermi surface on a finite
lattice even in this method. The divergence now comes from the line integral
equivalent to the one marked $1$ in Figure ~\ref{fig:cont}.  On the lattice 
one can perform the integral over $\omega$ to obtain the zero temperature 
part of the number density. Taking successive derivatives of this term with 
respect to $\mu$ gives the zero temperature artifacts in the linear $\mu$ method 
for the second and fourth order susceptibilities~\cite{gs12}:
\begin{equation}
\chi_{20}(T=0)=-\frac{1}{4 N^3}\sum_{\vec p}\left(1- \sqrt{\frac{f}{1+f}}\right)~,~
\chi_{40}(T=0)= -\frac{3}{4 N^3}\sum_{\vec p}\left(2-\frac{3+2f}{1+f} \sqrt{\frac{f}{1+f}}\right)~.
\label{eqn:susclm}
\end{equation}

\section{Testing the idea on the lattice }
\begin{figure}[htb]
\includegraphics[scale=0.6]{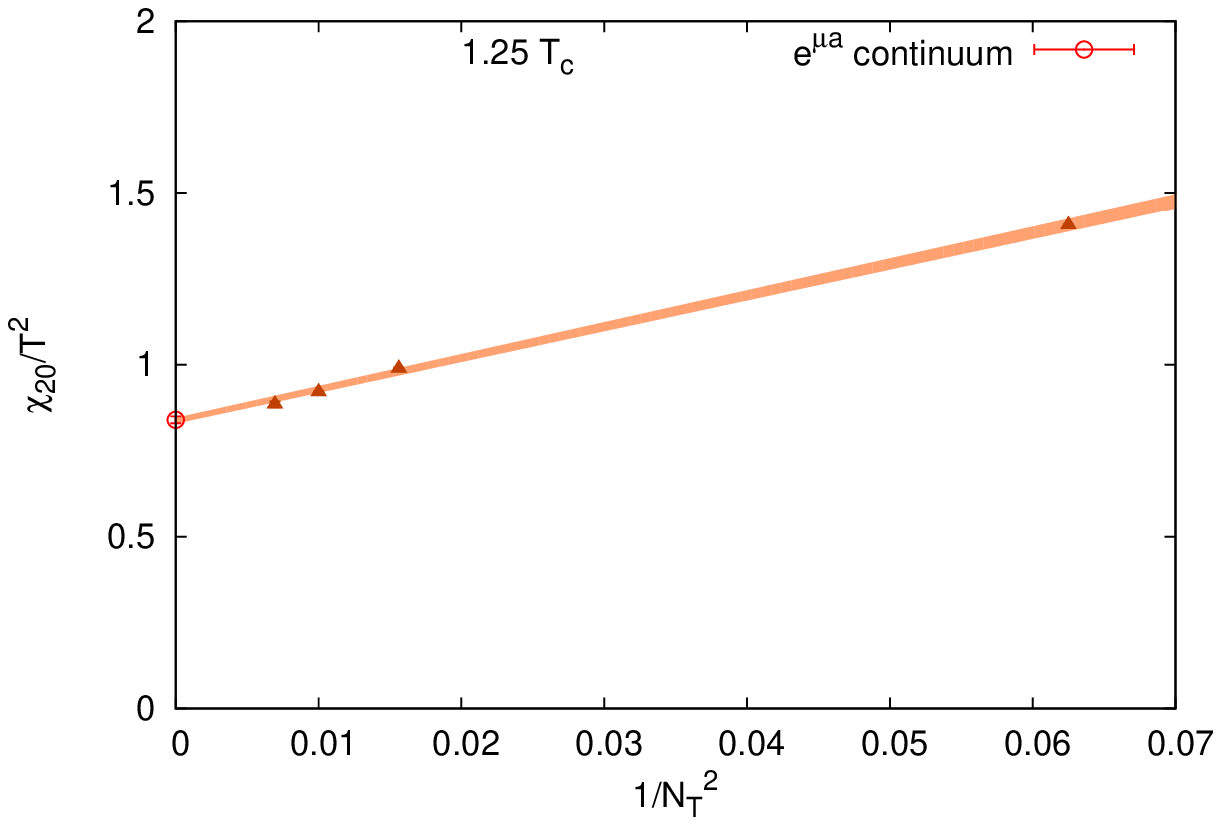}
\includegraphics[scale=0.6]{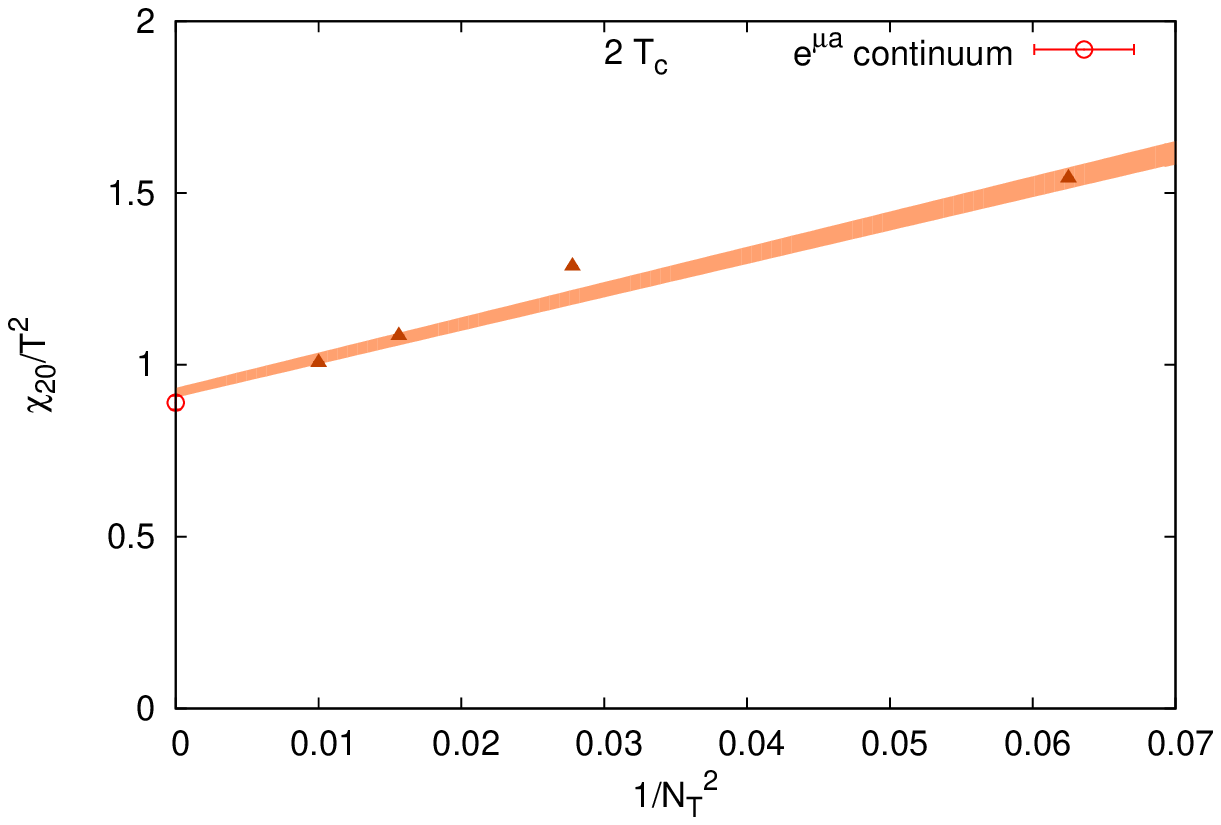}
\caption{The continuum extrapolated results for the second order quark number
susceptibilities in quenched QCD for a quark mass $m/T_c=0.1$ at
$1.25 T_c$ (left) and $2 T_c$ (right) compared to the corresponding
results using the $\rm{e}^{\mu a}$ method.} 
\label{fig:susc2}
\end{figure}

\begin{figure}[htb]
\includegraphics[scale=0.6]{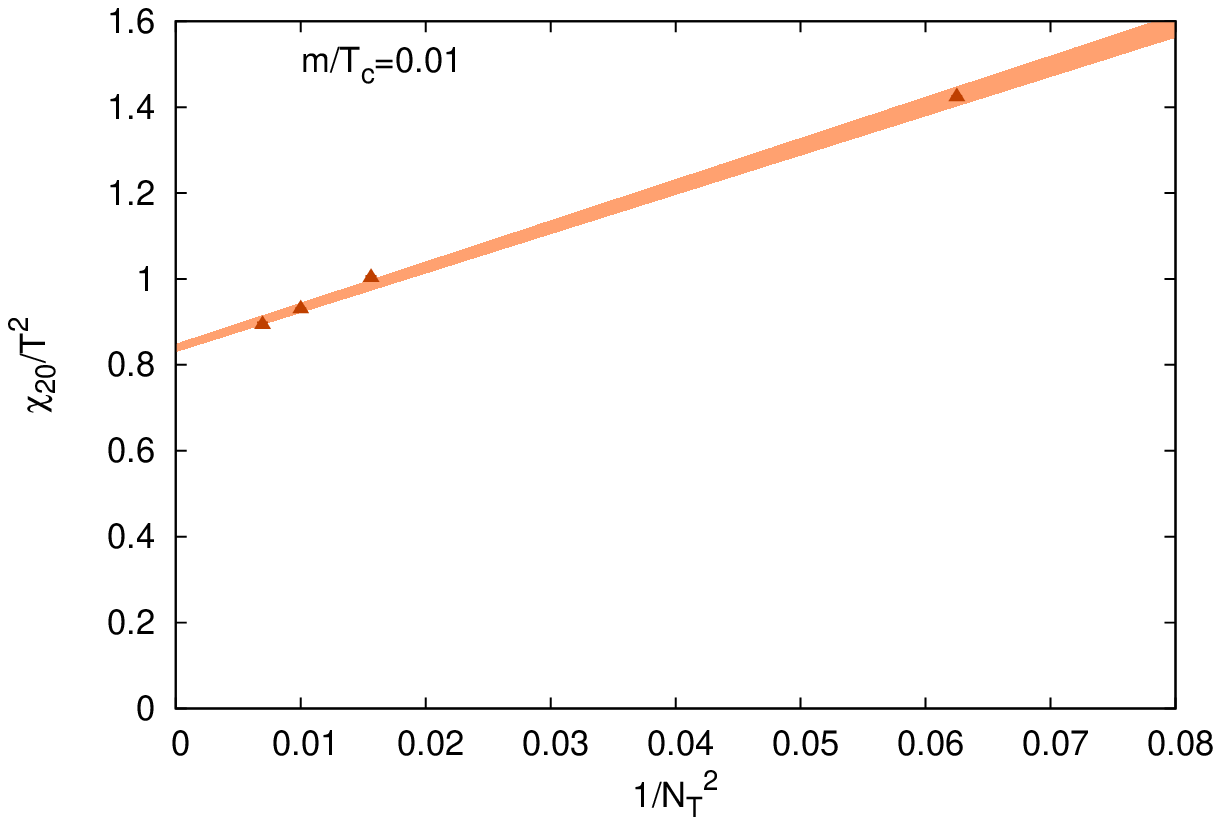}
\includegraphics[scale=0.6]{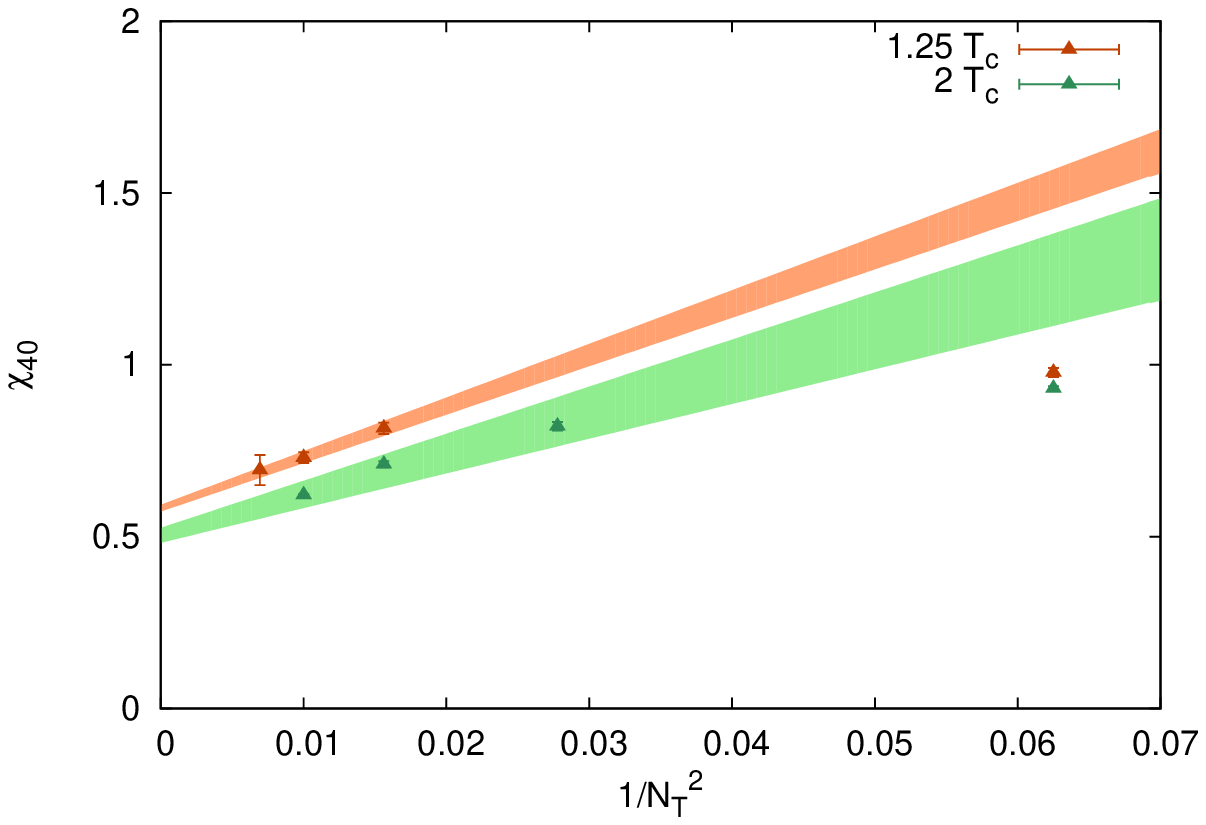}
\caption{The continuum extrapolated results for the second order quark number susceptibilities for a lighter quark mass $m/T_c=0.01$  
at $1.25 T_c$ (left) and the results for the fourth order susceptibility for
the quark mass $m/T_c=0.1$ (right). } 
\label{fig:susc4}
\end{figure}

Having noted that the divergence on the lattice is no different that a
$\Lambda$-cut-off in the continuum theory, we proposed, and demonstrated,
earlier that a subtraction of the $T=0$ ideal gas term for the quark number
susceptibility on the same size $N^3$ lattice does eliminate the divergence in
the continuum limit obtained by sending $N_T \to \infty$ for the free case~\cite{gs12}.
Here we concern ourselves with the test of that idea for interacting theory.  We
calculated the susceptibilities(QNS) for two degenerate quark flavours on
quenched lattices with the linear $\mu$ Dirac operator~\cite{gs14}. The pure
$SU(3)$ gauge configurations were generated using the Cabibbo-Marinari
pseudo-heatbath algorithm with three $SU(2)$ subgroup update per sweep using the
Kennedy-Pendleton updating method. We generated $N_T = 4, 6, 8, 10$ and $12$
lattices and $25$-$100$ independent configurations each at two different
temperatures given by $T/T_c=1.25, 2$.  To keep the finite volume effects under
control we checked and used $N_s=4 N_T$. The second order QNS was calculated by
inverting the Dirac operator on $400$ and $500$ Gaussian random vectors at $2,
1.25 T_c$ respectively.  The results obtained using this method contain
unphysical $\mu^2/a^2$ and $\mu^0$ artifacts for second and fourth order QNS.
For removing them we calculate the corresponding free theory QNS at zero
temperature in Eq. (\ref{eqn:susclm}) and perform the continuum
extrapolation. The expressions for these $T=0$ values and the corresponding
numerical values are 
displayed in Table \ref{table:1}. 
\begin{table}
 \centering
 \begin{tabular}{|c|c|c|l|l|}
 \hline \hline
 $N_T$ & $T$(MeV) & $m/T_c$ & $\chi_{20}(T=0)$ & $\chi_{40}(T=0)$  \\
 \hline
 4  &      &     &-0.062487   &-0.142923\\
 8  & 1.25 &     &-0.062499  &-0.142661\\
 10  &      & 0.1 &-0.062501   &-0.142620\\
 12 &      &     &-0.062502   &-0.142607\\
 \hline
 4  &   &     & -0.062541  &-0.142999\\
 6  & 2 &     & -0.062500  &-0.142611\\
 8  &   & 0.1 & -0.062508  &-0.142677\\
 10 &   &     & -0.062512  &-0.142718 \\
 \hline
  \end{tabular}
  \caption{The $T=0$ ideal gas subtraction terms at different volumes and 
temperatures for the second and fourth order QNS used in this work.}
  \label{table:1}
\end{table}
Performing the subtraction, the physical values of the second order QNS with a quark mass $m/T_c=0.1$ 
is shown in Figure~\ref{fig:susc2}. We do not observe any divergent term as evident from the 
positive slope of the data. Moreover, our extrapolated continuum result coincides with the earlier 
result obtained with the $\exp(\pm a \mu)$ action \cite{ggquen,swagato}, marked with red open circles on 
the vertical axis of each figure. In other to check that there are no mass dependent divergent terms 
we calculated the second order QNS with a still lower quark mass $m/T_c=0.01$, the continuum extrapolated 
results shown in the left panel of Figure~\ref{fig:susc4}. We indeed confirm absence of any new divergences 
from the slope of the fitted curve. Furthermore, removing the free theory $T=0$ artifacts 
from the fourth order QNS also gives the correct continuum limit as evident from the right panel of 
Figure~\ref{fig:susc4}.
For the QNS $\chi_n~,n\geq 6$, the additional artifacts in the linear $\mu$ method are $\mathcal{O}(a^{n-4})$, hence 
do not affect the continuum extrapolation of the lattice results. Moreover these terms reduce the cut-off effects compared 
to the $\rm{e}^\mu a$-method and facilitate a smoother approach to the continuum~\cite{gs12}.

\section{Summary}
We did a comprehensive study of the possible origin of the $\mu^2/a^2$ divergences in the quark number susceptibility 
on the lattice when we introduce chemical potential as a $\mu N$ term in the fermion action, $N$ being the conserved 
number density on the lattice. We found out that such a divergence explicitly exist for the QNS of free fermions with 
a cut-off regulator. The lattice regulator just faithfully reproduces this divergence for the free fermions as well. 
Conserved charges are not renormalized.   Assuming therefore that the only
divergence that would exist for QCD on the lattice is due to the $\mu N$ term of
the free theory, we did an explicit check in quenched QCD by taking the
continuum limit of the second and the fourth order ideal gas subtracted QNS. We
indeed find that  there
are no additional divergences that arise due to interactions. The
divergence in the linear $\mu$-method can be removed easily which gives us
another path towards introducing $\mu$ on the lattice in addition to the most
popular method~\cite{hk} where one modifies the quark action at finite $\mu$ to
remove these divergences explicitly. This method may be beneficial for
calculating the higher order QNS with considerably less computational effort
allowing us to calculate the Lattice EoS at finite baryon chemical potential and
make progress towards measuring the critical end-point.


\begin{thebibliography}{99}



\bibitem{gertcpod}
G.\ Aarts, Proceeding of this conference~.

\bibitem{hk}
P.\ Hasenfratz and F.\ Karsch  {\sl Phys.\ Lett.\  } B125, 308 (1983);
J.\ Kogut {\sl et al.}, {\sl Nucl.\ Phys.\ } B 225, 93 (1983).

\bibitem{bilgav}
N.\  Bilic and R.\ V. Gavai {\sl Z.\ Phys.\  } C23, 77 (1984).

\bibitem{gavai}
R.\ V.\ Gavai, {\sl Phys.\ Rev.\  } D32, 519 (1985).

\bibitem{ggquen}
R.\ V.\ Gavai and S.\ Gupta, {\sl Phys.\ Rev.\ } D67, 034501 (2003).

\bibitem{ggplb}
R.\ V.\ Gavai and S.\ Gupta,  {\sl Phys.\ Lett.\  } B696, 459 (2011).

\bibitem{gs10}
R.\ V.\ Gavai and S.\ Sharma, {\sl Phys.\ Rev.\  } D81, 034501 (2010).

\bibitem{piswilc}
 R.\ D.\ Pisarski and F.\ Wilczek, {\sl Phys.\ Rev.\ } D29, 338 (1984).
 
\bibitem{neunar}
        R.\ Narayanan and H.\ Neuberger,  {\sl Phys.\ Rev.\ Lett.\ } 71, 3251 (1993); \\
        H.\ Neuberger,  {\sl Phys.\ Lett.\ }  B417, 141 (1998).
        
\bibitem{kaplan}
        D.\ Kaplan,  {\sl Phys.\ Lett.\ } B288, 342 (1992).
                

\bibitem{hln} 
P.\ Hasenfratz, V.\ Laliena and F.\  Neidermeyer, 
 {\sl Phys.\ Lett.\ }  B427,   125 (1998). 

 \bibitem{Lues} 
M.\ Luscher,  {\sl Phys.\ Lett.\ } B428, 342 (1998).

\bibitem{dweos}
T.\ Bhattacharya et. al., {\sl Phys.\ Rev.\ Lett.\ } 113,  082001 (2014).

\bibitem{wettig}
J.\ Bloch and T.\ Wettig, {\sl Phys.\ Rev.\ Lett.\ } 97, 012003 (2006);\\
J.\ Bloch and T.\ Wettig, {\sl Phys.\ Rev.\  } D76, 114511 (2007).

\bibitem{gl} C.\ Gattringer and L.\ Liptak, {\sl Phys.\ Rev.\ } D76, 054502 (2007).

\bibitem{bgs} 
D.\ Banerjee, R.\ V.\ Gavai and S.\ Sharma, 
{\sl Phys.\ Rev.\ } D78,  014506 (2008);\\
D.\ Banerjee, R.\ V.\ Gavai and S.\ Sharma, {\sl PoS (LATTICE 2008)}, 177.

\bibitem{gsov}
R.\ V.\ Gavai and S.\ Sharma,  {\sl Phys.\ Lett.\ } B {\bf 716}, 446 (2012).

\bibitem{ns}
R.\ Narayanan and S.\ Sharma, {\sl JHEP\ } {\bf 1110}, 151 (2011);

\bibitem{gs12}
R.\ V.\ Gavai and S.\ Sharma, {\sl Phys.\ Rev.\  } D85, 054508 (2012).

\bibitem{gs14}
R.\ V.\ Gavai and S.\ Sharma, {\tt arXiv:1406.0474} [hep-lat].

\bibitem{gg1}
R.\ V.\ Gavai and S.\ Gupta, {\sl Phys.\ Rev.\  } D78, 114503 (2008).


\bibitem{swagato}
S.\ Mukherjee, {\sl Phys.\ Rev.\  } D74, 054508 (2006).











\end{thebibliography}
\end{document}